\documentclass{iopart}
\usepackage{epsfig}
\usepackage{subfigure}

\begin{document}
 
\article[Temperature dependent sound velocity in hydrodynamic equations ]
  {Strangeness in Quark Matter, Levo\v ca, Slovakia, June 2007}
  {Temperature dependent sound velocity in hydrodynamic equations for relativistic heavy-ion collisions 
  \footnote{Research supported by the Polish Ministry of Science and Higher Education grants N202~153~32/4247 and 
  N202~034~32/0918 for the years 2007-2009}
}  
 
\author{Miko{\l }aj Chojnacki}
\address{The H. Niewodnicza\'nski Institute of Nuclear Physics,\\
  Polish Academy of Sciences,\\
  ul. Radzikowskiego 152, PL-31342 Krak\'ow, Poland}
\ead{Mikolaj.Chojnacki@ifj.edu.pl}
   
\begin{abstract}
We analyze the effects of different forms of the sound-velocity function $c_s(T)$ on the hydrodynamic evolution of matter formed in the central region of relativistic heavy-ion collisions. At high temperatures (above the critical temperature $T_c$) the sound velocity is calculated from the recent lattice simulations of QCD, while in the low temperature region it is obtained from the hadron gas model. In the intermediate region we use different interpolations characterized by the values of the sound velocity at the local maximum (at $T = 0.4 \,T_c$) and local minimum (at $T=T_c$). In all considered cases the temperature dependent sound velocity functions yield the entropy density, which is consistent with the lattice QCD simulations at high temperature. Our calculations show that the presence of a distinct minimum of the sound velocity leads to a very long ($\sim$ 20 fm/c) evolution time of the system, which is not compatible with the recent estimates based on the HBT interferometry. Hence, we conclude that the hydrodynamic description is favored in the case where the cross-over phase transition renders the smooth sound velocity function with a possible shallow minimum at $T_c$.
\end{abstract}
\submitto{\JPG}
\maketitle

The recently analysed RHIC data suggests that matter formed in relativistic heavy-ion collisions behaves like a perfect fluid \cite{Heinz:2005zg}. This hypothesis triggers new developments in relativistic hydrodynamics \cite{Teaney:2003kp,Hirano:2005xf,Hama:2005dz,Eskola:2005ue,Heinz:2005bw,Huovinen:2006jp,Andrade:2006yh,Baier:2006um,Koide:2006ef,Nonaka:2006yn,Satarov:2006jq,Hirano:2007xd,Bialas:2007gn}. 
In this paper we continue the studies of Ref. \cite{Chojnacki:2007jc} and discuss the effects of different forms of the sound-velocity function $c_s(T)$ on the hydrodynamic evolution of matter formed in the central region of relativistic heavy-ion collisions.

\begin{figure}[ht!]
\begin{center}
\subfigure{\includegraphics[angle=0,width=0.49\textwidth]{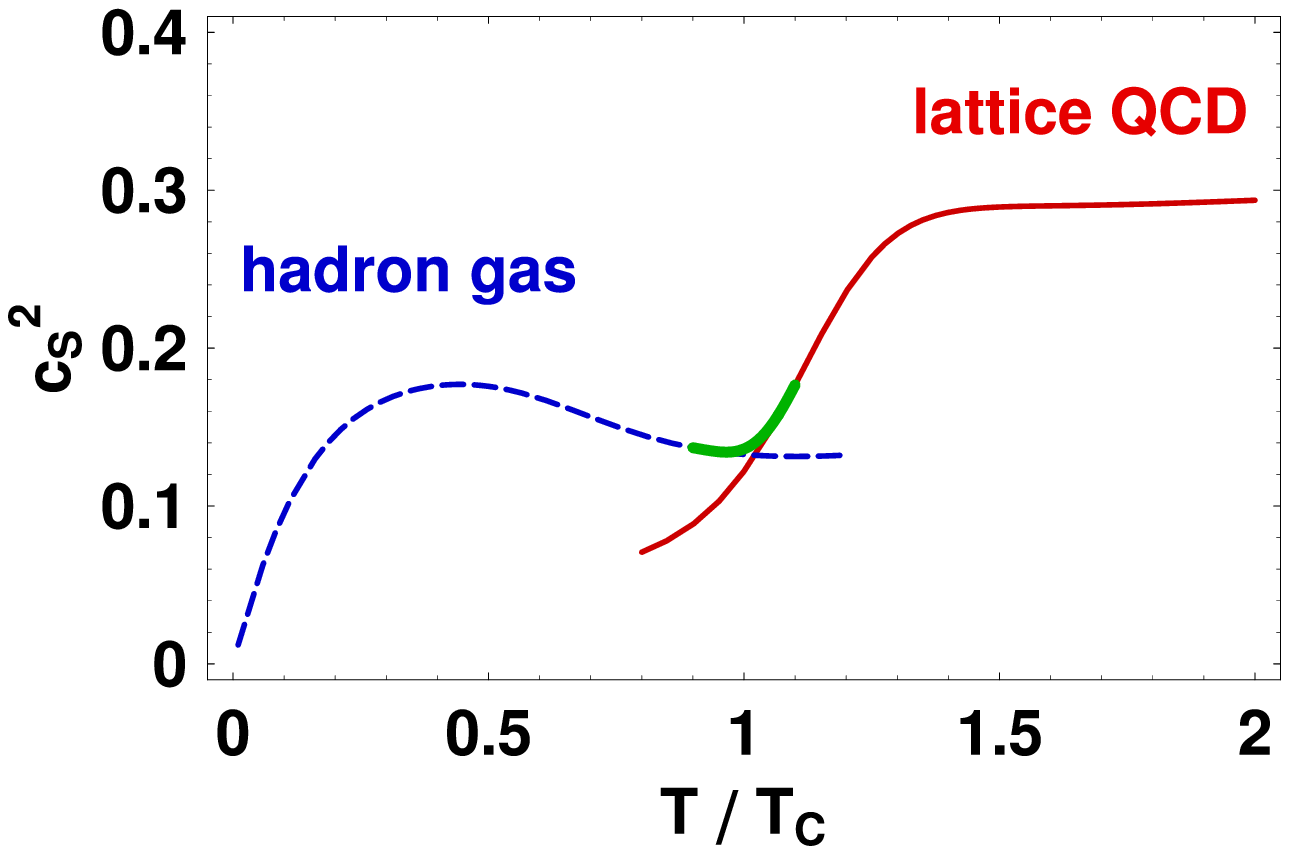}}
\subfigure{\includegraphics[angle=0,width=0.49\textwidth]{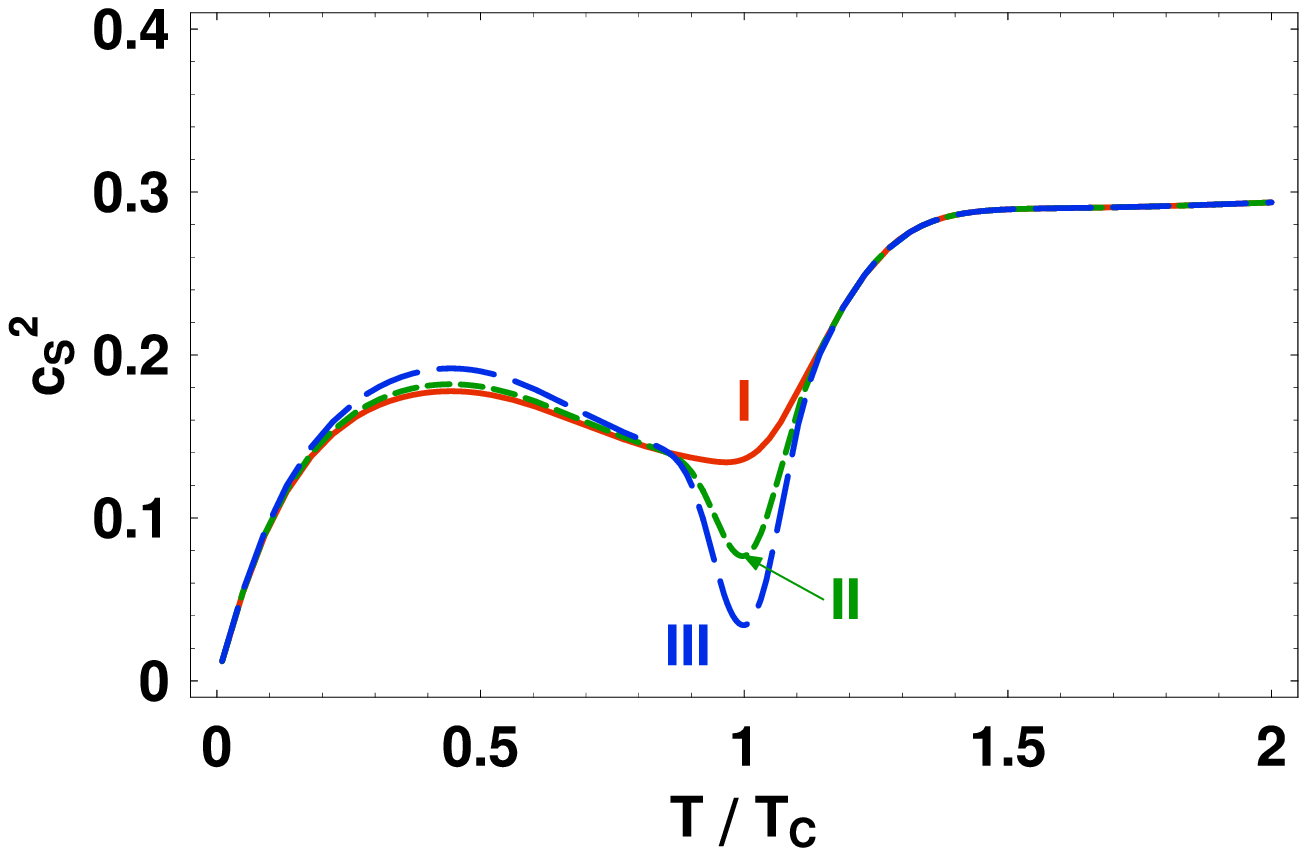}}
\end{center}
\caption{The sound velocity squared, $c_s^2(T)$, shown as a function of the temperature at zero baryon density. The left panel shows the results of the lattice simulations of QCD \cite{Aoki:2005vt} (solid line) and the ideal hadron gas model \cite{Chojnacki:2004ec} (dashed line). The simplest interpolation between the two calculations is marked by a thick solid line. The right panel presents the three cases of the function $c_s(T)$ considered in our calculation. The case I agrees with the interpolation shown in the left panel. A deeper local minimum at $T_c$ is introduced in the case II (dashed line) and III (long-dashed line) to model different forms of the phase transition. The sound velocity at the critical temperature is 25\% (case II) and 50\% (case III) lower in comparison to the case I. A corresponding increase of the sound velocity around $T \approx 0.4 \,T_c$ in the cases II and III is required to have the same entropy density at high temperatures. The increase of the sound velocity in this region may account for pourly known repulsive van der Waals forces in the hadron gas.}
\label{fig:cs2HGQCD}
\end{figure}

\begin{figure*}[ht!]
\begin{center}
\includegraphics[angle=0,width=0.6\textwidth]{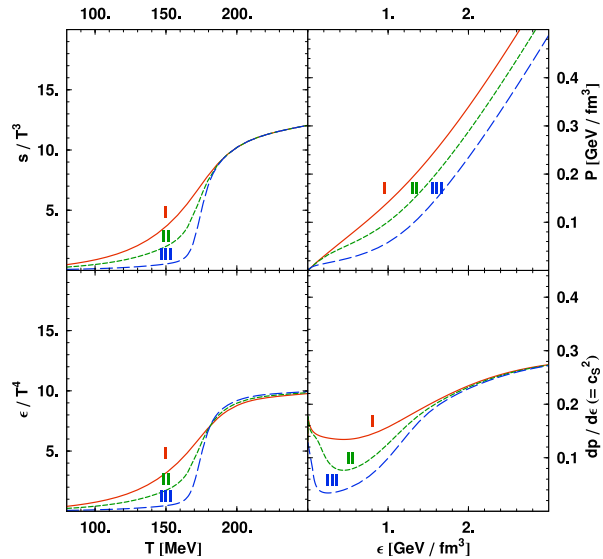}
\end{center}
\caption{The temperature dependence of the entropy density and energy density, the two upper panels, as well as the energy density dependence of the pressure and sound velocity, the two lower panels. One can observe that the deeper is the minimum of the sound velocity function, the steeper is the increase of the entropy density and the energy density.}
\label{fig:huo}
\end{figure*}

The left panel of Fig. \ref{fig:cs2HGQCD} shows the sound velocity function $c_s^2(T)$ obtained from the lattice simulations of QCD \cite{Aoki:2005vt} (solid line)
and the ideal hadron gas model \cite{Chojnacki:2004ec} (dashed line). The lattice results were obtained for physical masses of the light quarks and the strange quark. One can observe that the sound velocity is below the limiting value  $c_s^2=\frac{1}{3}$ expected for a non-interacting gas of massless particles. On the other hand, in the low temperature part the hadron-gas calculation is used which includes full mass spectrum of resonances \cite{Broniowski:2000bj}. It is interesting to check that for very small temperatures ($T << m_\pi$) the sound velocity is determined by the properties of a massive pion gas giving $c_s(T) = \sqrt{T/m_\pi}$, where $m_\pi$ is the mass of the pion. The lattice and hadron-gas calculations overlap in the region close to the critical temperature $T_c$ = 170 MeV and can be naturally joined by an interpolating function (thick solid line). A smooth function like that would result in a cross-over phase transition. A different way to connect the two results is to introduce a local minimum at the critical temperature. As the value of the $c_s$ decreases, the phase transition more and more resembles the first order phase transition where, by definition, the sound velocity drops to zero at $T = T_c$ . 

Since the lattice QCD simulations are not very much reliable in the region $T < T_c$ and the hadron-gas calculations are meaningless in the region $T > T_c$ we decided to consider different interpolations between the lattice and hadron-gas results in the vicinity of the phase transition. In this respect our study may be regarded as supplementary to Ref. \cite{huovinen2005} where the impact of different equations of state on the particle spectra and $v_2$ was studied. The three considered by us interpolations are denoted below as the case I, II and III, for details see Fig. \ref{fig:cs2HGQCD}. All considered functions are consistent with the lattice simulations of QCD for the temperatures $T > 1.15 \,T_c$ and with the ideal hadron gas for the $T < 0.15 \,T_c$. In the temperature range $0.85 \,T_c < T < 1.15 \,T_c$ the three cases differ by the depth of the local minimum. The value of the sound velocity at the critical temperature $T_c$ equals $0.37$ in the case I. In the case II it is decreased by 25\% compared to the case I (\mbox{$c_s(T_c)=0.28$}) and in the case III it is decreased by 50\% (\mbox{$c_s(T_c)=0.19$}). In order to be able to compare the solutions obtained for such three different cases, it is crucial to have the same entropy density at high temperatures, where the lattice data are reliable. To achieve this requirement we use the following thermodynamic relation connecting the entropy density $s(T)$ with the sound velocity for zero baryon chemical potential
\begin{equation}
s(T) = s(T_0) \exp\left[ \,\,\,\int\limits_{T_0}^T \,\frac{dT^\prime}{T^\prime c_s^2(T^\prime)}
\right].
\label{sumrule}
\end{equation}
One may conclude from Eq. (\ref{sumrule}) that the decrease of the sound velocity results in the relative increase of the entropy density in the high temperature region. To avoid the entropy increase we increase the values of the sound velocity in the region $0.15 \,T_c < T < 0.85 \,T_c$. This kind of modification may be interpreted as an effective parametrization of the repulsive van der Waals forces in the hadron gas. In Fig. \ref{fig:huo} we present several thermodynamic functions calculated for the 3 considered sound velocity functions. The upper left plot shows the functions $s(T)/T^3$ where, as required, all the lines converge at high temperatures. All presented sound velocity functions have been also checked to fulfill the condition against the formation of shock waves \cite{Baym:1983sr,Blaizot:1987cc}

A detailed explanation of the method to solve the relativistic hydrodynamic equations with the temperature dependent sound velocity may be found in Ref. \cite{Baym:1983sr} and in our later developments of this approach in  \cite{Chojnacki:2004ec,Chojnacki:2006tv}, see also 
\cite{Dyrek:1984xz}. When considering RHIC experiments in the midrapidity region one can assume with good justification boost-invariance and zero baryon chemical potential \cite{Florkowski:2001fp,Torrieri:2004zz}. With this simplification the only independent thermodynamic parameter is the temperature and the equations of relativistic hydrodynamics have the following form
\begin{eqnarray}
u^{\mu }\partial _{\mu }\left( T\,u^{\nu }\right) =\partial ^{\nu }T, \quad
\partial _{\mu }\left(s u^{\mu }\right) =0,
\label{acc}
\end{eqnarray}
where $T$ is the temperature, $s(T)$ is the entropy density, and $u^{\mu}=\gamma \left( 1,\mathbf{v}\right) $ is the hydrodynamic four-velocity. Further transformations include introduction of the temperature dependent sound velocity $c_s \left( T\right)$ and the potential $\Phi \left( T\right)$ according to the  formulas
\begin{equation}
c_{s}^{2}=\frac{\partial P}{\partial \varepsilon }=\frac{s}{T}\frac{\partial T}{\partial s },\qquad d\Phi =\frac{d\ln T}{c_{s}}=c_{s}d\ln s.
\label{cs}
\end{equation}
The relativistic boost-invariant hydrodynamic equations with cylindrical symmetry may be expressed in the compact form with the help of the auxiliary functions $A_+$ and $A_-$ defined as
\begin{equation}
A_{\pm } = \Phi \pm \alpha,
\label{apm}
\end{equation}
where $\alpha$ is the transverse rapidity of the fluid element, $v_{r}=\tanh\alpha$. The longitudinal velocity in the considered case has the well known form $v_{z}=z/t$ \cite{Bjorken:1983qr}  and Eqs. (\ref{acc}) read
\begin{eqnarray}
\frac{\partial A_{\pm }\left( t,r\right) }{\partial t} +\frac{v_{r}\pm c_{s}} {1\pm v_{r}\,c_{s}}\frac{\partial A_{\pm }\left( t,r\right) }{\partial r}\, + \frac{c_{s}}{1\pm v_{r}\,c_{s}}\left( \frac{v_{r}}{r}+\frac{1}{t}\right) = 0.
\label{eqapm}
\end{eqnarray}
All interesting physical observables can be calculated from the auxiliary functions $A_\pm$ with the explicit formulas given in Refs. \cite{Chojnacki:2007jc,Chojnacki:2004ec,Chojnacki:2006tv}.


For symmetry reasons, both the velocity field $v_r$ and the temperature gradient $\partial T/\partial r$ should vanish at $r=0$. This condition is achieved if the functions $A_+$ and $A_-$ are initially determined by a single function $A(r)$ according to the prescription  \cite{Baym:1983sr}
\begin{eqnarray}
A_+(t=t_0,r)=A(r), \quad A_-(t=t_0,r)=A(-r).
\label{inita}
\end{eqnarray}
Our main physical assumption about the initial state is that the initial temperature profile is connected with the nucleon-nucleus thickness function $T_A(r)$ by the following equation
\begin{equation}
T(t_0,r) =  T_s\left[ s(t_0,r) \right]   =
T_s\left[ s_0 \frac{T_A(r)}{T_A(0)} \right]   .
\label{T01}
\end{equation}
Here $T_s(s) $ is the inverse function to the entropy density function $s(T)$ and the parameter $s_0$ is the initial entropy at the center of the system. The idea to use Eq. (\ref{T01}) follows from the assumption that the initially produced entropy density  $s(t_0,r)$ is proportional to the density of wounded nucleons at a distance $r$ from the collision center. We use the value $s_0=70.5 \, \hbox{fm}^{-3}$ which yields \mbox{$T(t=t_0,r=0) = 2 T_c$}. 
We note that the functions $s(T)$ and $T_s(s)$ (evaluated for the cases I, II, and III) agree if the temperature or entropy is sufficiently large. Hence, in the three considered cases the initial temperature profiles are practically the same in the center of the system. The small differences appear however if we consider larger values of $r$ where the temperature and entropy drops down. We recall that the thickness function is defined by the equation
\begin{equation}
T_A(r) = 2 \int dz \, \rho\left(\sqrt{r^2+z^2}\right),
\label{TA}
\end{equation}
where the function $\rho(r)$ is the nuclear density profile given by the Woods-Saxon function with a conventional choice of the parameters: $\rho_0 = 0.17 \,\hbox{fm} ^{-3}$, \mbox{$r_0 = (1.12 A^{1/3} -0.86 A^{-1/3})$ fm}, $\, a = 0.54 \,\hbox{fm}, \, A = 197$. 
We note that the initial condition (\ref{T01}) may be included in the initial form of the function $A(r)$ 
\begin{equation}
A(t=t_0,r) = \Phi_T \left\{
T_s\left[ s_0 \frac{T_A(r)}{T_A(0)} \right] \right\} . 
\label{initaT}
\end{equation}

\begin{figure}[t]
\begin{center}
\subfigure{\includegraphics[angle=0,width=0.49\textwidth]{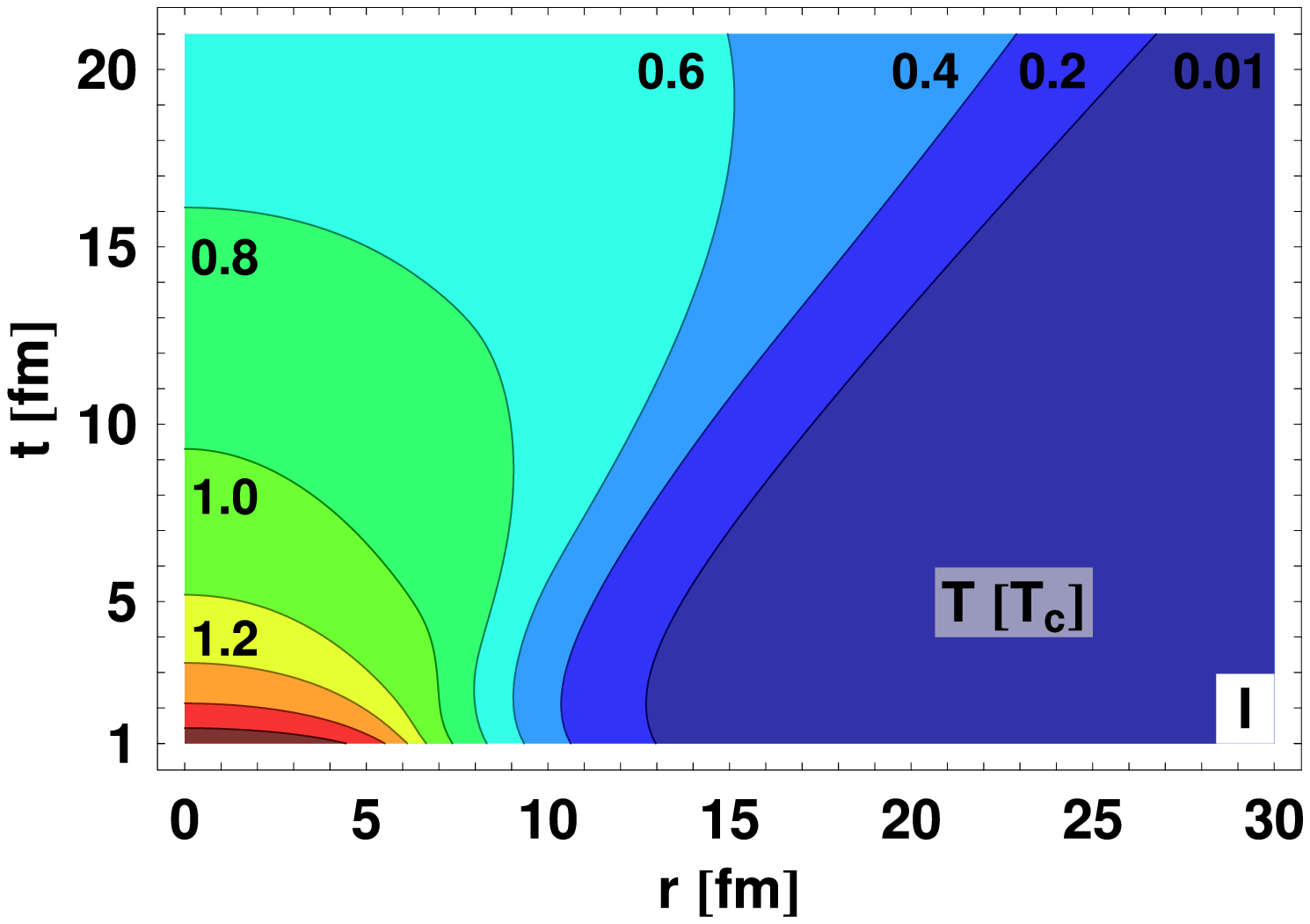}}\\
\subfigure{\includegraphics[angle=0,width=0.49\textwidth]{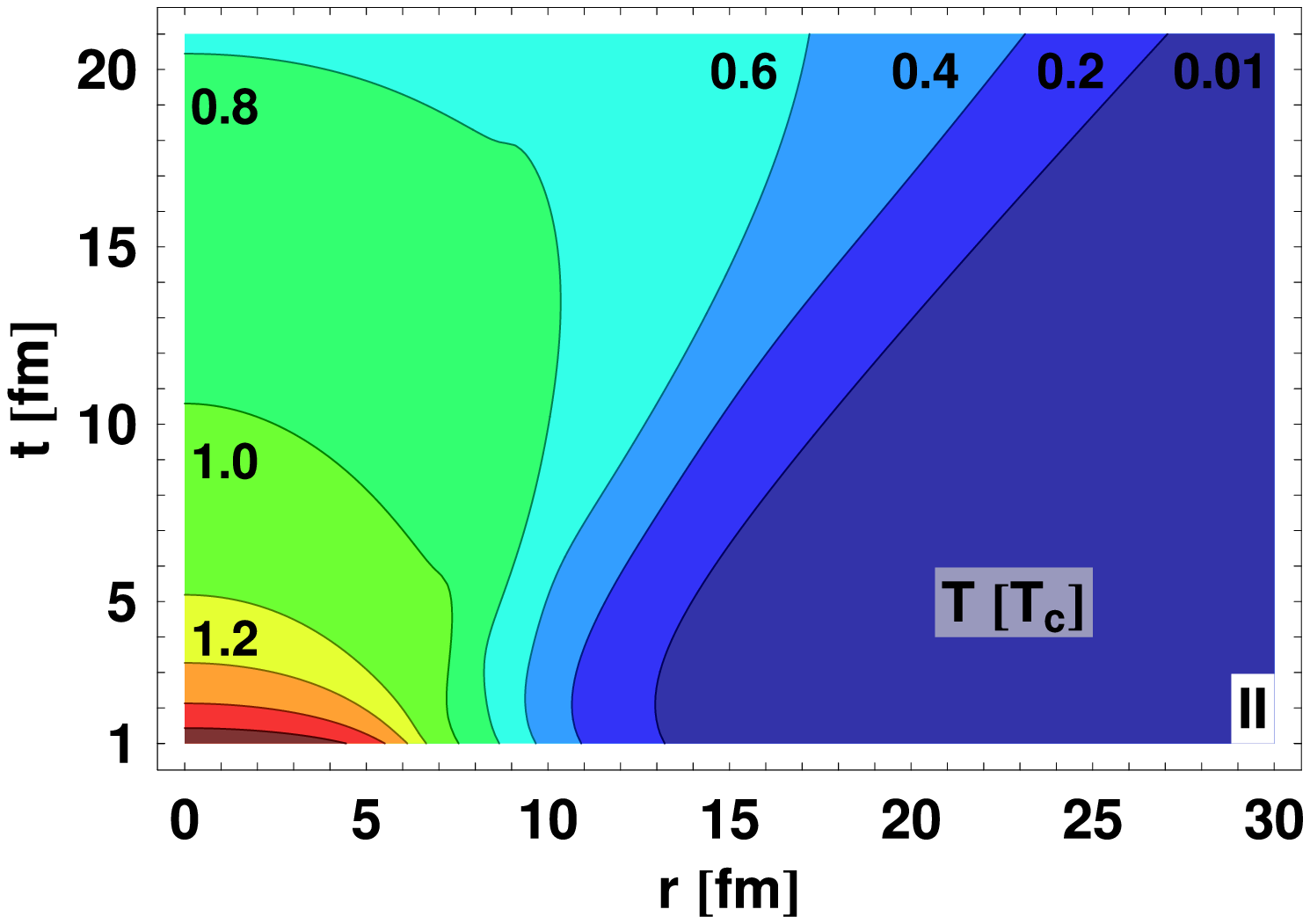}}
\subfigure{\includegraphics[angle=0,width=0.49\textwidth]{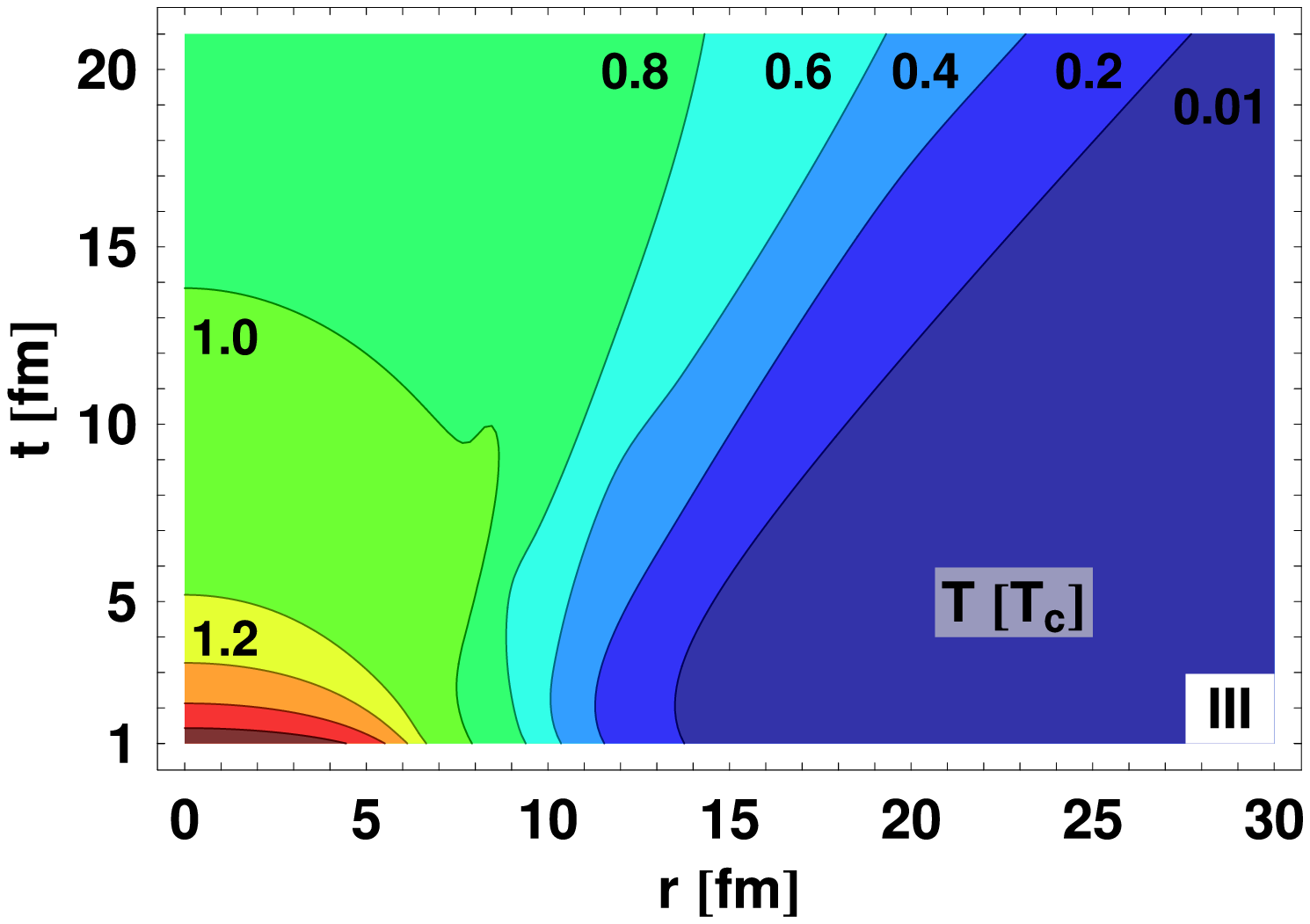}}
\end{center}
\caption{Isotherms describing the hydrodynamic evolution for the case I; sound velocity with a shallow minimum where $c_s(T_c)$ = 0.37; case II sound velocity with a moderate minimum, $c_s(T_c)$ = 0.28 and case III with a deep minimum, \mbox{$c_s(T_c)$ = 0.19}. The numbers denote the values of the temperature in the units of the critical temperature $T_c$ = 170 MeV.}
\label{fig:TIsodip}
\end{figure}

In Fig. \ref{fig:TIsodip} we show the isotherms describing the hydrodynamic evolution of the system in the cases I, II and III. The numbers at the isotherms give the values of the temperature in the units of the critical temperature $T_c$. In the case I, the upper panel, we observe that the center of the system cools down to 0.8 $T_c$ after the evolution time of about 15 fm. We note that the hydrodynamic description should be replaced (around $T \sim 0.8 \, T_c$) by the model describing hadronic rescattering (see, e.g., Ref. \cite{Nonaka:2006yn}) whose presence additionally increases the lifetime of the system. Another option is to assume that freeze-out happens at high temperature (see Refs. \cite{Broniowski:2001we,Broniowski:2002nf}) where many physical observables were successfully reproduced under the assumption of a universal freeze-out taking place at the temperature of 165 MeV). In the latter case the lifetime of the system may be identified simply with the time when the system passes the phase transition. In the discussed case this time is of about 10 fm.

In the lower left panel of Fig. \ref{fig:TIsodip} we show the isotherms describing the hydrodynamic evolution of the system with the sound velocity II. The initial entropy is exactly the same as in the case I. One can notice that the central temperature does not drop below 0.8 $T_c$ before 20 fm. Clearly, the dip in the sound velocity causes a dramatic increase of the lifetime of the system. The most striking situation is presented in the lower right part of Fig. \ref{fig:TIsodip} where the sound velocity with the deepest minimum is considered. One can notice that the system does not pass the phase transition before the considered evolution time of 20 fm. Contrary, even at that time the system has the tendency to expand more, the effect indicated by the shapes of the isotherms.  

The observation of the effects related with the possible presence of the softest point of the QCD equation of state triggered the ideas about the long-living states formed in ultra-relativistic heavy-ion collisions \cite{Hung:1995eq,Rischke:1995cm,Heinz:2005ja}. In view of the RHIC data indicating a short lifetime of the system we may analyze the problem of the lifetime in the reverse order, asking the question how soft the equation of the state may be to allow for a hydrodynamic description consistent with the HBT results. Our results show that the smooth behavior of the sound velocity without any distinct minimum is favored if we demand the short hydrodynamic evolution time of about 10 fm.

\end{document}